\documentclass[conference]{IEEEtran}
\usepackage{cite}
\usepackage{amsmath,amssymb,amsfonts}
\usepackage{amssymb}
\usepackage{algpseudocode}
\usepackage{amsfonts}
\usepackage{graphicx}
\usepackage{fancyhdr}
\usepackage{cases}
\usepackage{textcomp}
\usepackage{extarrows}
\usepackage{algorithm,algpseudocode}
\usepackage{multirow}
\usepackage{graphicx}
\usepackage{overpic}
\usepackage{graphicx}
\usepackage{booktabs}
\usepackage[caption=false,font=footnotesize]{subfig}
\IEEEoverridecommandlockouts

\usepackage{orcidlink}
\hypersetup{hidelinks} % hide the colorful frame line
\usepackage{multirow,tabularx}
\usepackage{mathtools}
\usepackage{xcolor}
\usepackage[english]{babel}
\usepackage{amsthm}

\IEEEoverridecommandlockouts
\def\BibTeX{{\rm B\kern-.05em{\sc i\kern-.025em b}\kern-.08em
    T\kern-.1667em\lower.7ex\hbox{E}\kern-.125emX}}
    
\begin{document}

\bstctlcite{BSTcontrol}
\title{Robust Continuous-Time Beam Tracking with Liquid Neural Network}
% \thanks{X. Wang, F. Zhu, Q. Zhou, Q. Yu and C. Huang are with the College of Information Science and Electronic Engineering, Zhejiang University, Hangzhou 310007, China (e-mail:chongwenhuang@zju.edu.cn).}
% \thanks{C. Yuen is with the School of Electrical and Electronics
% Engineering, Nanyang Technological University, Singapore. }

% \thanks{M. Debbah is with Khalifa University of Science and Technology, P O Box  127788, Abu Dhabi, UAE (email: merouane.debbah@ku.ac.ae).}
\author{
    \IEEEauthorblockN{Fenghao Zhu$^{1, 2}$, Xinquan Wang$^{1}$, Chongwen Huang$^{1, 2}$, Richeng Jin$^{1}$, Qianqian Yang$^{1}$, Ahmed Al Hammadi$^{3}$, \\Zhaoyang Zhang$^{1}$,  Chau Yuen$^{4}$,~\IEEEmembership{Fellow,~IEEE}, and M\'{e}rouane~Debbah$^{5}$,~\IEEEmembership{Fellow,~IEEE}
    }
    \IEEEauthorblockA{$^1$ College of Information Science and Electronic Engineering, Zhejiang University, Hangzhou 310027, China}
    \IEEEauthorblockA{$^2$ State Key Laboratory of Integrated Service Networks, Xidian University, Xi’an 710071, China}
    \IEEEauthorblockA{$^3$ Technology Innovation Institute, 9639 Masdar City, Abu Dhabi, UAE}
    \IEEEauthorblockA{$^4$ School of Electrical and Electronics Engineering, Nanyang Technological University, Singapore 639798 }
    \IEEEauthorblockA{$^5$ KU 6G Research Center, Khalifa University of Science and Technology, P O Box 127788, Abu Dhabi, UAE}
    % \IEEEauthorblockA{$^5$ CentraleSupelec, University Paris-Saclay, 91192 Gif-sur-Yvette, France}
    }

\maketitle

\begin{abstract}
Millimeter-wave (mmWave) technology is increasingly recognized as a pivotal technology of the sixth-generation communication networks due to the large amounts of available spectrum at high frequencies. However, the huge overhead associated with beam training imposes a significant challenge in mmWave communications, particularly in urban environments with high background noise. To reduce this high overhead, we propose a novel solution for robust continuous-time beam tracking with liquid neural network, which dynamically adjust the narrow mmWave beams to ensure real-time beam alignment with mobile users. Through extensive simulations, we validate the effectiveness of our proposed method and demonstrate its superiority over existing state-of-the-art deep-learning-based approaches. Specifically, our scheme achieves at most 46.9\% higher normalized spectral efficiency than the baselines when the user is moving at 5 m/s, demonstrating the potential of liquid neural networks to enhance mmWave mobile communication performance.
\end{abstract}


\section{Introduction}\label{sec:intro}

\addtolength{\topmargin}{-.178in}

Millimeter-wave (mmWave) communication is a promising enabling technology for improving the data rate of wireless communications due to its abundance of spectrum resources at high frequencies \cite{6g, 6gsurvey, 6G_Science_China}. To fully exploit this potential, massive multiple-input multiple-output (MIMO) configurations have been deployed in both base stations (BSs) and user equipment (UE) to increase system throughput \cite{tang2023hybrid}. However, the high-dimensional signal processing associated with massive MIMO imposes a significant computational overhead \cite{va2016beam}. Additionally, the presence of background noise in urban environments with complex electromagnetic landscapes poses a substantial challenge in determining the optimal beam directions \cite{gmml}.
\par
To mitigate this challenge, beam tracking techniques have been employed to reduce the overhead of beam training by predicting future beam directions. In \cite{EKF}, a second-order model was presented for estimating angle-of-departure (AoD) and angle-of-arrival (AoA) variations with an extended Kalman filter-based beam tracking framework. However, this model assumed slow variations in angular velocity and did not account for scenarios with highly mobile users or dynamic changes in angular velocity.
\par
Machine learning (ML) is adept at leveraging historical data to address new problems with similar data structures \cite{chinacom_AI_6G, wbhVTC}. Previous researches have demonstrated that ML-based techniques can effectively perform beamforming with mmWave \cite{zhu2023robust, wgan, WangGML, zhu2024robust}. Specifically, long short-term memory (LSTM) neural networks have been employed to predict beams based on past pilot signals \cite{lstm1, lstm2}.  Through nonlinear fitting to model the features of UE movements, the variations in AoA and AoD can be more precisely captured, resulting in improved performance compared to conventional methods. However, due to the inherent design of LSTM for discrete-time processing, it falls short in real-time beam adjustments \cite{sherstinsky2020fundamentals}. To overcome this limitation, \cite{lstm-ode} integrated the LSTM structure with ordinary differential equation (ODE) to predict the optimal beam at arbitrary instants. By incorporating ODEs to model the derivative of beam variations, this approach achieved higher beamforming gains in continuous-time beam tracking tasks.
\par
Nonetheless, the above approaches overlooked the deployment of beam tracking in urban areas characterized by intricate electromagnetic landscapes, where users are frequently moving amidst substantial interference noise. Traditional deep learning methods exhibited limited robustness in such noisy settings, leading to a significant degradation in performance under heavy interference \cite{ncps_beamforming}. To tackle the challenges posed by dynamic scenarios with noise, a specialized type of recurrent neural network called liquid neural network (LNN) has been introduced \cite{LNN-AAAI}. Inspired by the nervous system of Caenorhabditis elegans, LNN emulates the time-varying behavior of synapses in the brain, which enables the network to excel at continuous-time data processing. It demonstrated exceptional adaptability in various applications such as autopilot systems \cite{lechner2020neural} and flight navigation \cite{chahine2023robust}. In this paper, we propose a robust continuous-time beam tracking scheme employing LNN. Specifically, the received pilot signal vectors undergo feature extraction, then the extracted features are processed by the LNN unit. Finally, the output layer produces a probability vector that determines the selection of the beam from the codebook. Leveraging the modeling of LNN, our proposed method can predict the sub-optimal beam at arbitrary instants. Furthermore, the flexibility of the nonlinear response in the designed model enables our scheme to adaptively process rapidly changing noisy channels. Simulation results confirm the effectiveness of our proposed method in achieving up to 46.9\% higher normalized spectral efficiency than the state-of-the-art baselines with the UE moving at 5 m/s.
\par
The rest of this paper is organized as follows: Section \ref{sec:sys} describes the system model and problem formulation. Section \ref{sec:method} provides a comprehensive analysis of the LNN-based beam tracking method, including the architecture of LNN and the flowchart of the proposed deep learning model. Simulation results are presented in Section \ref{sec:simulation} to verify the performance of the proposed algorithm. Finally, the conclusions are outlined in Section \ref{sec:conclusion}.

\vspace{4mm}
\section{System Model and Problem Formulation}\label{sec:sys}

\addtolength{\topmargin}{-.10in}

\begin{figure}[t]\vspace{0mm}
	\begin{center}
		\centerline{\includegraphics[width=0.5\textwidth]{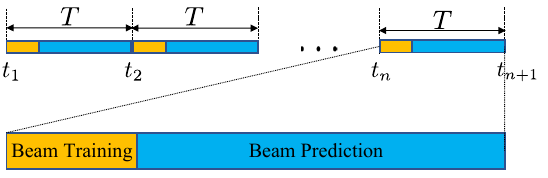}}  
		\vspace{-0mm}
		\captionsetup{name={Fig.}, labelsep=period}
		\caption{The beam tracking diagram.}
		\label{fig:tracking_diagram} \vspace{-4mm}
	\end{center}
\end{figure}

Consider a downlink mmWave communication system with a $N_t$-antenna BS and a single-antenna user. We denote the transmitted symbol as $x$ which satisfies $E\{ |x| \}=1$. Regarding the channel model, we adopt the narrowband block-fading channel model from the DeepMIMO dataset \cite{alkhateeb2019deepmimo}. The channel vector $\mathbf{h}_{d}$ comprises $L$ channel paths, each with a time delay $\tau_{l} \in \mathbb{R}$ and azimuth/elevation AoA $\theta_{l}, \varphi_{l}$, respectively. $\rho$ denotes the path loss between the user and the BS, $\alpha_l$ is the complex gain, and $p(\tau_l)$ represents a pulse shaping function for $T_{S}$-spaced signalling evaluated at time $\tau_l$. Therefore, the channel vector can be formulated as
\begin{equation}\label{DeepMIMO Channel Model}
\mathbf h_{d}= \sqrt \frac{N_t}{\rho}\sum_{l=1}^{L}\alpha_{l} p(dT_{S}-\tau_{l})  \mathbf a(\theta_{l}, \varphi_{l}),
\end{equation}
where $\mathbf a(\theta_{l}, \varphi_{l})$ is the array response vector of the BS at the AoA $\theta_{l}, \varphi_{l}$.
% To quantify the channel estimation error (CEE), we refer to the perfect channel as $\mathbf{h}$ and the estimated channel as $\hat{\mathbf{h}}$. Then the CEE can be measured in decibels (dB) as follows
% \begin{equation}\label{CEE}
%     \mathrm{CEE} = 10\log_{10}\left(\frac{\mathbb{E}[\| \mathbf{h} -  \hat{\mathbf{h}} \|_2^2]}{\mathbb{E}[\| \mathbf{h} \|_2^2]}\right).
% \end{equation}
% CEE serves as a metric for evaluating the accuracy of channel estimation. A smaller value of CEE indicates a more precise channel estimation.
\par
We assume that the BS implements analog beamforming using phase shifters with a single radio frequency chain. The discrete Fourier transform (DFT) codebook $\mathcal{M}$ comprises ${Q}$ candidate beams. The $q$-th beam $\mathbf{v}^{(q)}$ in $\mathcal{M}$ can be represented as
\begin{equation}\label{codebook}
    \mathbf{v}^{(q)} = \sqrt{\frac{1}{N_t}}[1, e^{j\frac{2\pi}{Q}q}, \dots, e^{j\frac{2\pi}{Q}(N_t-1)q}]^T.
\end{equation}
The received signal $y$ at the user, considering the channel vector $\mathbf{h}$, can be expressed as
\begin{equation}\label{received signal}
    y = \sqrt{P}\mathbf{h}^H\mathbf{v}x + n,
\end{equation}
where $P$ is the transmit power and $n$ is the additive noise following a circularly symmetric complex Gaussian distribution with zero mean and covariance $\sigma^{2}$. The signal-to-noise ratio (SNR) is defined as $\mathrm{SNR} = P / \sigma^2$. Therefore, the spectral efficiency of the system can be written as 
\begin{equation}\label{SE}
    R(\mathbf{h}, \mathbf{v}) = \log_2(1 + \mathrm{SNR}\cdot |\mathbf{h}^H\mathbf{v}|^2).
\end{equation}
The aim of this paper is to maximize the target function as described in \eqref{SE} by choosing the best beam in $\mathcal{M}$. Therefore, the classification problem can be expressed as
\begin{equation}\label{optimization problem}
    \mathbf{v}^{*} = \mathop{\mathrm{argmax}}_{q \in \{1,2,\dots, Q\}}R(\mathbf{{h}}, \mathbf{v}^{(q)}).
\end{equation}

\vspace{-0mm}

We assume that $T$ is the length of a time slot, as illustrated in Fig. \ref{fig:tracking_diagram}. Each time slot comprises two stages: beam training and beam prediction. We define the received signal vector of the $n$-th beam training as $\mathbf{y}_n = [y_n^{(1)}, y_n^{(2)},\dots,y_n^{(Q)}]^T$ and the past received signal matrix of the previous $n$ beam trainings as $\mathbf{Y}_n = [\mathbf{y}_1, \mathbf{y}_2,\dots,\mathbf{y}_n]$. During the beam training stage, conventional beam training is conducted to obtain the optimal beam using the received signal vector $\mathbf{y}_n$. Subsequently, during the beam prediction stage, knowledge from the previously received signal matrix $\mathbf{Y}_n$ is utilized to continuously predict the optimal beam. Let $t$ denote the arbitrary time between two consecutive beam training instants $t_n$ and $t_{n+1}$. Then the normalized instant $\overline{t}$ can be defined as
\begin{equation}\label{normalized time}
    \overline{t} = \frac{t-t_n}{T} \in [0, 1],
\end{equation}
 The choice of beam $\hat{q}(t_{n} + \overline{t}T)$ at the normalized instant $\overline{t}$ can be formulated as
\begin{align}\label{beam choice}
     \hat{q}(t_{n} & + \overline{t}T) = \mathbf{S}(\mathbf{Y}_n, \overline{t}), \\
     \mathrm{s.t.\ \ } & \hat{q} \in \{1,2,\dots, Q\}, \\
     & \overline{t} \in [0, 1],
\end{align}
where $\mathbf{S(\cdot)}$ denotes the classification method. 
% It can be implemented by either traditional analytical methods or the latest deep learning methods.

\vspace{4mm}
\section{LNN-based Beam Tracking}\label{sec:method}
In this section, we first introduce the synapse dynamics and the architecture of LNN. Then the continuous-time beam tracking model based on LNN is presented.

\subsection{LNN Architecture}
\begin{figure}[t]\vspace{0mm}
	\begin{center}
		\centerline{\includegraphics[width=0.5\textwidth]{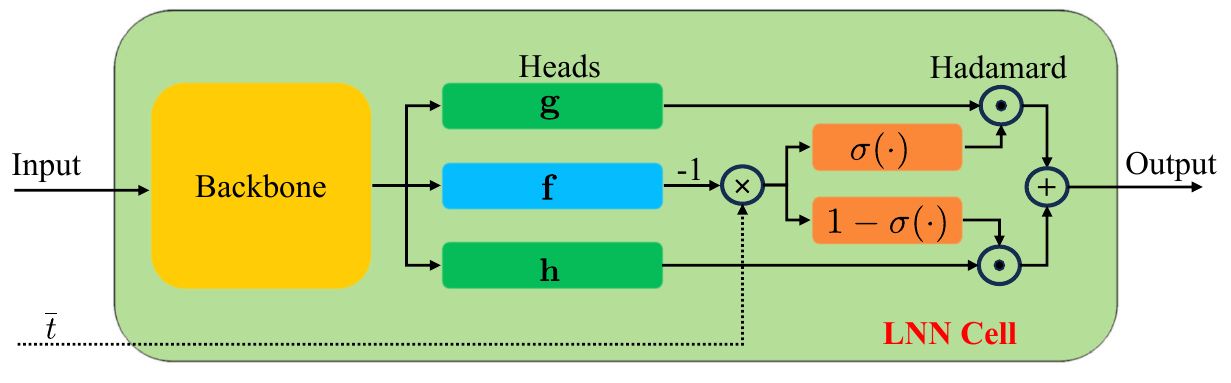}}  
		\vspace{-0mm}
		\captionsetup{name={Fig.}, labelsep=period}
		\caption{The LNN cell architecture. Here, $\sigma(\cdot)$ refers to the sigmoid activation function.}
		\label{fig:LNN_Architecture} \vspace{-4mm}
	\end{center}
\end{figure}
\vspace{-0mm}
The fundamental model of neural and synapse dynamics \cite{lechner2020neural} is described as follows: The stimulation $i(t) \in \mathbb{R}$ is transmitted to a postsynaptic neuron via a nonlinear conductance-based synapse model. The membrane potential of the postsynaptic neuron is denoted as $x(t) \in R$, which can be formulated by the following ODE
\begin{align}\label{synapse dynamics}
    &  \frac{\mathrm{d}x(t)}{\mathrm{d}t} = -\frac{x(t)}{\tau} + s(t), \\
    &  s(t) = f(i(t))(a-x(t)),
\end{align}
where $s(t)$ represents the nonlinear synapse, $f(\cdot)$ represents the non-linear synaptic release, while $a$ and $\tau$ stand for the synaptic reversal potential and the time constant respectively. Considering the interaction of neurons, we can vectorize \eqref{synapse dynamics} as follows
\begin{align}\label{synapse dynamics vectorize}
       \frac{\mathrm{d}\mathbf{x}(t)}{\mathrm{d}t} = & -[\mathbf{\omega}_{\tau} + \mathbf{f}(\mathbf{x}(t), \mathbf{i}(t); \mathbf{\theta})] \odot \mathbf{x}(t) \\   & + \mathbf{a} \odot \mathbf{f}(\mathbf{x}(t), \mathbf{i}(t); \mathbf{\theta}) \nonumber.
\end{align}
Here, from the perspective of neural networks, $\mathbf{x}(t) \in \mathbb{R}^{(D \times 1)} $ denotes the hidden states of all neurons. $\mathbf{\omega}_{\tau}$ represents the time constant parameter vector. $\mathbf{i}(t) \in \mathbb{R}^{(m \times 1)} $ denotes the aggregated input from adjacent neurons. $\mathbf{a}$ is a bias vector, and $\mathbf{f}(\cdot)$ represents the neural network parameterized by $\mathbf{\theta}$. Finally, $\odot$ is the Hadamard product operator.
\par
The equation \eqref{synapse dynamics vectorize} serves as a fundamental building block of the LNN. However, solving ODE-based problems typically involves iterative methods, which can be computationally intensive. Recently, a closed-form expression has been introduced to approximate the solution with lower complexity \cite{hasani2022closed}, expressed as follows
\begin{align}\label{closed-form}
    \mathbf{x}(t) = & \underbrace{\sigma(-\mathbf{f}(\mathbf{x}(t), \mathbf{i}(t); \theta_{\mathbf{f}}){t})}_{\mathrm{time-continuous\, gating}} \odot \mathbf{g}(\mathbf{x}(t), \mathbf{i}(t); \mathbf{\theta}_{\mathbf{g}}) \\ + & \underbrace{[\vec{\mathbf{1}} - \sigma(-\mathbf{f}(\mathbf{x}(t), \mathbf{i}(t); \mathbf{\theta}_{\mathbf{f}}){t})]}_{\mathrm{time-continuous\, gating}} \odot \mathbf{h}(\mathbf{x}(t), \mathbf{i}(t); \theta_{\mathbf{h}}), \nonumber
\end{align}
where $\mathbf{f}(\cdot)$ acts as the liquid time constant for the sigmoid gates, while $\mathbf{g}(\cdot)$ and $\mathbf{h}(\cdot)$ construct the non-linearity of the LNN. Instead of modelling $\mathbf{f}(\cdot)$, $\mathbf{g}(\cdot)$ and $\mathbf{h}(\cdot)$ separately with three neural networks, we adopt a shared architecture approach. Specifically, we employ a feedforward neural network named backbone, which is followed by a Tanh activation layer. Subsequently, the network branches into three distinct single-layer feedforward neural networks, each referred to as the head neural network. This design facilitates the learning of the coupling effect between the liquid time constant and non-linearity components, while still retaining a degree of independence. Fig. \ref{fig:LNN_Architecture} illustrates the fundamental architecture of the LNN cell. At the normalized time instant $\overline{t}$, the input $\mathbf{x}$ is concatenated with the previous hidden states. This concatenated input is then fed into the LNN cell, which processes the data and outputs the approximate hidden states of the neurons as described in equation \eqref{synapse dynamics vectorize}. Thanks to the powerful feature extraction and modelling ability of neural networks, the closed-form expression \eqref{closed-form} can effectively approximate the solution of \eqref{synapse dynamics vectorize} after sufficient training. This circumvents the high computational overhead associated with traditional ODE solvers, which rely on iterative methods for convergence.
\begin{figure}[t]\vspace{0mm}
	\begin{center}
		\centerline{\includegraphics[width=0.5\textwidth]{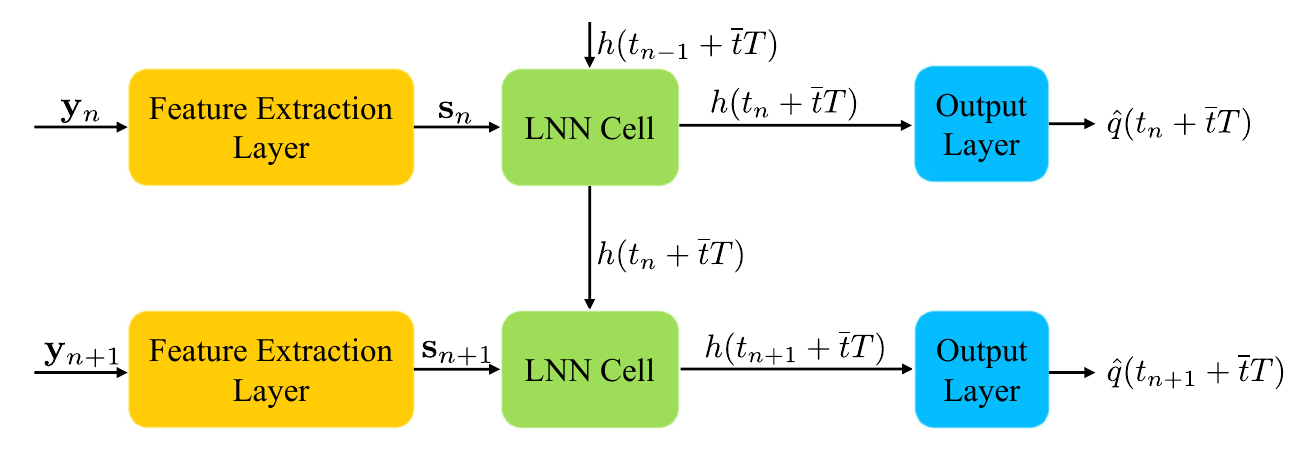}}  
		\vspace{-0mm}
		\captionsetup{name={Fig.}, labelsep=period}
		\caption{Flowchart of proposed deep learning model based on LNN.}
		\label{fig:workflow} \vspace{-2mm}
	\end{center}
\end{figure}
\vspace{-0mm}

\begin{algorithm}[t]
\caption{LNN-based Beam Tracking}
\label{alg:lnn_algorithm}
\begin{algorithmic}[1]
    \State \textbf{Input:} Time slot $T$, received signal vector $\mathbf{y}_n$ during beam training and the corresponding time instant
    $t_n$, the beam prediction instant $t \in [t_n, t_{n+1}]$.
    \State Initialize the weights of the feature extraction layer, LNN cell and the output layer.
    \State $\overline{t} = (t-t_n)/T$
    \For{$i\leftarrow 1,2,\cdots,n$}
        \State $\mathbf{s}_i = \mathrm{FeatureExtraction}(\mathbf{y}_i)$
        \State $h(t_{i} + \overline{t}T) = \mathrm{LNNCell}(\mathbf{s}_i, h(t_{i-1} + \overline{t}T), \overline{t})$
        \State $\hat{\mathbf{p}}(t_{i} + \overline{t}T) = \mathrm{OutputLayer}(h(t_{i} + \overline{t}T))$
    \EndFor
    \State $\hat{q}^*(t_{n} + \overline{t}T) = \mathop{\mathrm{argmax}}_{q \in \{1,2,\dots, Q\}}\hat{p}_q(t)$
    % \State $\mathrm{Loss}(t_{n} + \overline{t}T) = -\sum_{q=1}^{Q} p_q(t)\log\hat{p}_q(t)$
    \State \Return $\hat{q}^*(t_{n} + \overline{t}T)$
\end{algorithmic}
\end{algorithm}

\subsection{Proposed Beam Tracking Model}

\addtolength{\topmargin}{-.095in}

During model training, the input for prediction consists of the received signal vector $\mathbf{y}_n$. The optimal beam indices of arbitrary instants are saved as the classification label, which can be obtained through convention beam training. Once the model is fully trained, beam training is no longer needed in the prediction stage. The flowchart of proposed deep learning model based on LNN is presented in Fig. \ref{fig:workflow}. The detailed process is illustrated as follows: the received signal vector $\mathbf{y}_n$ from the $n$-th beam training slot is divided into real and imaginary parts, which are concatenated and fed into the feature extraction layer. The feature extraction layer consists of four batch normalization layers and three convolutional layers with the rectified linear unit (ReLU) activation function, which can extract the preliminary features $\mathbf{s}_n$ for subsequent processing, written as follows
\begin{equation}\label{gennerate feature}
    \mathbf{s}_n = \mathrm{FeatureExtraction}(\mathbf{y}_n).
\end{equation}
Afterwards, the LNN cell generates the new hidden states using the previous hidden states, input feature $\mathbf{s}_n$, and the normalized arbitrary instant $\overline{t}$, as depicted by the equation
\begin{equation}\label{gennerate hidden state}
    h(t_{n} + \overline{t}T) = \mathrm{LNNCell}(\mathbf{s}_n, h(t_{n-1} + \overline{t}T), \overline{t}).
\end{equation}
Subsequently, the generated hidden states in the $n$-th beam training slot is sent into the output layer, which comprises a fully connected layer followed by a Softmax layer. The output layer normalizes the hidden states into a probability vector, denoted as
\begin{equation}\label{softmax classification}
    \hat{\mathbf{p}}(t_{n} + \overline{t}T) = \mathrm{OutputLayer}(h(t_{n} + \overline{t}T)),
\end{equation}
where $\hat{\mathbf{p}}(t_{n} + \overline{t}T) = \hat{\mathbf{p}}(t) = [\hat{p}_1(t), \hat{p}_2(t), \dots, \hat{p}_Q(t)]$ and $\hat{p}_q(t)$ is the probability of the $q$-th candidate beam. The beam with the maximum probability is chosen for beam tracking, and the index is expressed as
\begin{equation}\label{max probability}
    \hat{q}^*(t_{n} + \overline{t}T) = \mathop{\mathrm{argmax}}_{q \in \{1,2,\dots, Q\}}\hat{p}_q(t).
\end{equation}
For this classification problem, we choose the Cross-Entropy loss function for model training, depicted as follows
\begin{equation}\label{loss function}
    \mathrm{Loss}(t_{n} + \overline{t}T) = -\sum_{q=1}^{Q} p_q(t)\log\hat{p}_q(t),
\end{equation}
where $p_q(t)=1$ indicates that the $q$-th beam is the optimal beam, otherwise $p_q(t)=0$. Algorithm \ref{alg:lnn_algorithm} summarizes the proposed beam tracking algorithm.

\begin{table}[t]\vspace{0mm}
\centering
\caption{Simulation Parameters}\vspace{-0mm}
\begin{tabular}{c cc c} 
\toprule
Parameters & Value & Parameters & Value\\ [0.2ex] 
\midrule
$ T_{total} $ & 1.6 s& Central Frequency & 28 GHz \\
$ T $ & 160 ms & Bandwidth & 50 MHz \\
$ Q $ & 64 & $N_\mathrm{F}$ & 9 dB \\
$ N_t $ &  64 & $P$ & 10 dBm \\
\bottomrule
\end{tabular}\vspace{-0mm}
\label{tab:simulation parameters}
\end{table}

\begin{table}[t] \vspace{-0mm}
    \centering
    \caption{Network Parameters}\vspace{-0mm}
    \label{Network_detail}
    \begin{tabular}{c c c c}
     \toprule
     No.&Layer Name &\emph{BF-Net} \\
     \midrule
     1& BN1 &$I=2, O=2$\\
     2& CNN1 & $I=2, O=64$, (3, 3, 1), ReLU  \\
     3& BN2  & $I=64, O=64$\\
     4& CNN2 & $I=64, O=256$, (3, 3, 1), ReLU\\
     5& BN3 & $I=64, O=256$ \\
     6& CNN3 & $I=256, O=256$, (3, 3, 1), ReLU\\
     7& BN4  & $I=256, O=256$ \\ 
     8& Pooling  & $I=256, O=256$, Average Pooling \\ 
     9& LNN Backbone  & $I=320, O=128$, Tanh \\ 
     10& LNN Head $\mathbf{f}$  & $I=128, O=64$ \\ 
     11& LNN Head $\mathbf{g}$  & $I=128, O=64$, Tanh \\ 
     10& LNN Head $\mathbf{h}$  & $I=128, O=64$, Tanh \\ 
     11& Output FC  & $I=64, O=64$, Softmax \\ 
     \bottomrule
    \end{tabular}\vspace{-0mm}
\end{table}

\vspace{4mm}
\section{Simulation Results}\label{sec:simulation}
In this section, we evaluate the performance of the proposed method in comparison to baseline approaches. We first illustrate the detailed system settings, followed by the presentation of the simulation results.
\subsection{System Settings}

\addtolength{\topmargin}{-.075in}

We investigate a mobile mmWave wireless communication system, and the DeepMIMO dataset \cite{alkhateeb2019deepmimo} is adopted to make the simulations more accurate and practical. The detailed simulation parameters are listed in Table \ref{tab:simulation parameters}. We select BS 1 of scenario 1 from the O1\_28 dataset for our experiments. The UE moves at speeds of $v \in \{5, 20\}$ m/s in random directions for each time slot, representing low and high-speed scenarios, respectively. The noise is defined as $\sigma^2 = (-174 + 10\log_{10}W + N_\mathrm{F})\,  \mathrm{dBm}$, where $W$ is the bandwidth and $N_F$ is the noise factor. The simulation consists of 10 time slots, each with a duration of 160 ms, resulting in a total beam tracking time of $T_{total}=1.6$ s. We aim to predict the optimal beams at arbitrary instant. For simplicity, we restrict the normalized instant $\overline t$ to the set $\{0.1, 0.2, \dots, 0.9\}$, which indicates we use the first 16 ms for beam training and predict the optimal beam every 16 ms in the rest of a time slot.
\par
The network parameters are detailed in Table \ref{Network_detail}, where $I/O$ represents the number of input or output channels or features. The tuple $(a, b,c)$ denotes the kernel size, sampling stride, and zero-padding size, respectively. The output FC refers to the fully connected layer with Softmax activation. We construct a dataset comprising 10,240 samples for training and 2,560 samples for validation. The model is trained for 100 epochs using the Adam optimizer with the learning rate and batch size set to $3 \times 10^{-5}$ and 32, respectively. To assess the performance, we utilize the normalized spectral efficiency as a metric, defined as follows:
\begin{equation}\label{normalized SE}
    \mathrm{SE_N} = \frac{R(\mathbf{{h}}, \mathbf{v}^{(\hat{q})})}{R(\mathbf{{h}}, \mathbf{v}^{(\hat{q}^*)})},
\end{equation}
where $R(\mathbf{{h}}, \mathbf{v}^{(\hat{q})})$ represents the current spectral efficiency, and $R(\mathbf{{h}}, \mathbf{v}^{(\hat{q}^*)})$ represents the optimal spectral efficiency. 
% Another metric, beam accuracy ($\mathrm{B_a}$), can be defined as:
% \begin{equation}
%     \mathrm{B_a} =  \frac{N_p}{N_p + N_n},
% \end{equation}
% where $N_p$ is the number of predicted beams that match the optimal beams, and $N_n$ is the number of incorrect beams.
We run the simulations on a computer equipped with an EPYC 75F3 CPU and an RTX 3090 GPU using PyTorch 2.0.1 and Python 3.9. As baselines, we employ the LSTM \cite{lstm1} and the LSTM-ODE \cite{lstm-ode}, whose feature extraction layer and output layer are identical to those of the proposed method in this paper. Since LSTM alone cannot handle continuous-time prediction, we assume that the LSTM baseline predicts the same beam for the beam prediction within a time slot.

\begin{figure}[t]\vspace{0mm}
	\begin{center}
		\centerline{\includegraphics[width=0.50\textwidth]{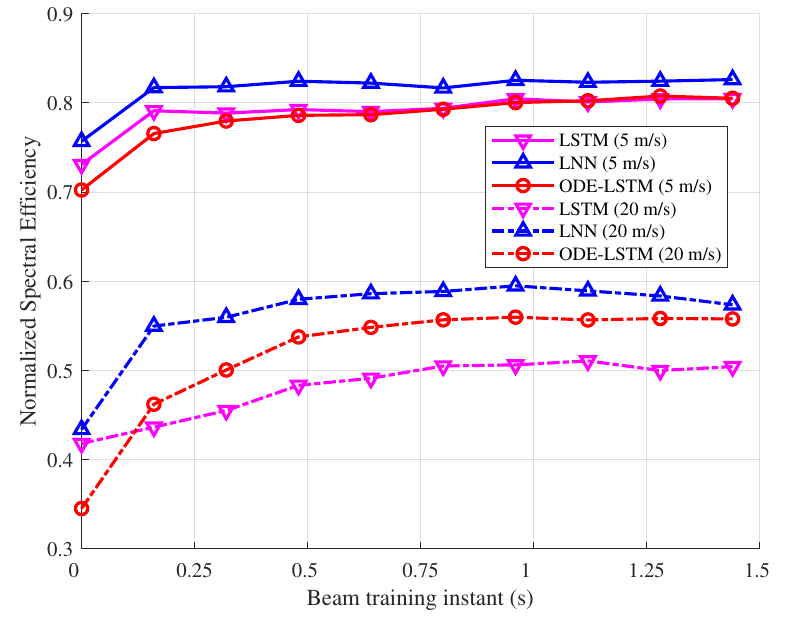}}  
		\vspace{-0mm}
		\captionsetup{name={Fig.}, labelsep=period}
		\caption{Performance evaluation with respect to beam training instant.}
		\label{fig:beam_training_time_instant} \vspace{-4mm}
	\end{center}
\end{figure}
\vspace{-0mm}

\subsection{Simulation Results}
Fig. \ref{fig:beam_training_time_instant} compares the normalized spectral efficiency $\mathrm{SE_N}$ of different schemes with respect to beam training instants, with UE velocities set to $v = 5$ and $20$ m/s.  The $\mathrm{SE_N}$ is averaged across all 9 normalized prediction instants within each of the 10 time slots. It is evident that all schemes exhibit an increase in $\mathrm{SE_N}$ with an increase in time. This can be attributed to the increased number of past received beam training signals, which help the neural networks learn better about the UE movement patterns. Notably, the proposed LNN-based method demonstrates superior performance compared to other schemes, particularly evident in the high-speed scenario ($v = 20$ m/s). Despite a decrease in $\mathrm{SE_N}$ as the UE speed increases from 5 m/s to 20 m/s, our proposed scheme maintains its superiority over the baselines, demonstrating robust performance over varying speeds.
\begin{figure}[t]\vspace{0mm}
	\begin{center}
		\centerline{\includegraphics[width=0.50\textwidth]{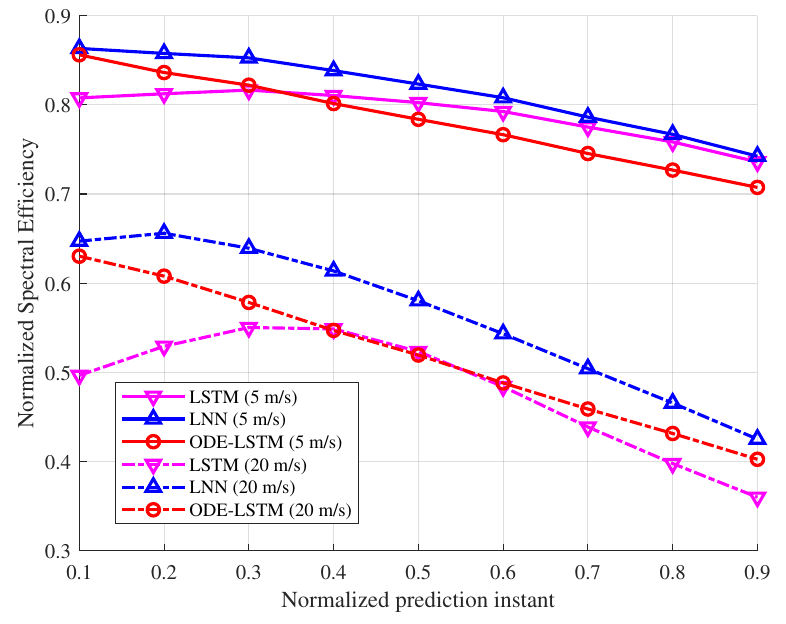}}  
		\vspace{-0mm}
		\captionsetup{name={Fig.}, labelsep=period}
		\caption{Performance evaluation with respect to normalized prediction instant.}
		\label{fig:normalized_prediction_instant} \vspace{-4mm}
	\end{center}
\end{figure}
\vspace{-0mm}

Fig. \ref{fig:normalized_prediction_instant} compares the normalized spectral efficiency $\mathrm{SE_N}$ of different schemes with respect to normalized prediction instants $\overline{t}$, considering UE velocities of $v = 5$ and $20$ m/s. The $\mathrm{SE_N}$ is averaged across all 10  time slots for each of the 9 normalized prediction instants. It is obvious that the performance of LSTM-ODE and the proposed scheme generally deteriorates with increasing normalized prediction instant $\overline{t}$, indicating 
rising uncertainty in UE movement since the previous beam training. Notably, the conventional LSTM scheme demonstrates an obvious single-peak performance in the high-speed scenario ($v = 20$ m/s). This is because we have assumed that the LSTM baseline predicts the same beam for the 9 beam predictions within a time slot, as LSTM can only predict the optimal beam for a single normalized instant. It would tend to accurately predict the middle-instant optimal beam to increase the overall performance. In contrast, ODE-LSTM and the proposed scheme LNN are both modelled on ODEs, which empower them to predict the optimal beam at any instant. However, the more accurate neuronal modelling of LNN facilitates better data variation capture and results in higher $\mathrm{SE_N}$. As the UE speed increases from 5 m/s to 20 m/s, the superior performance of LNN over other schemes becomes more pronounced, implying its suitability for high-speed applications.

\begin{figure}[t]\vspace{0mm}
	\begin{center}
		\centerline{\includegraphics[width=0.50\textwidth]{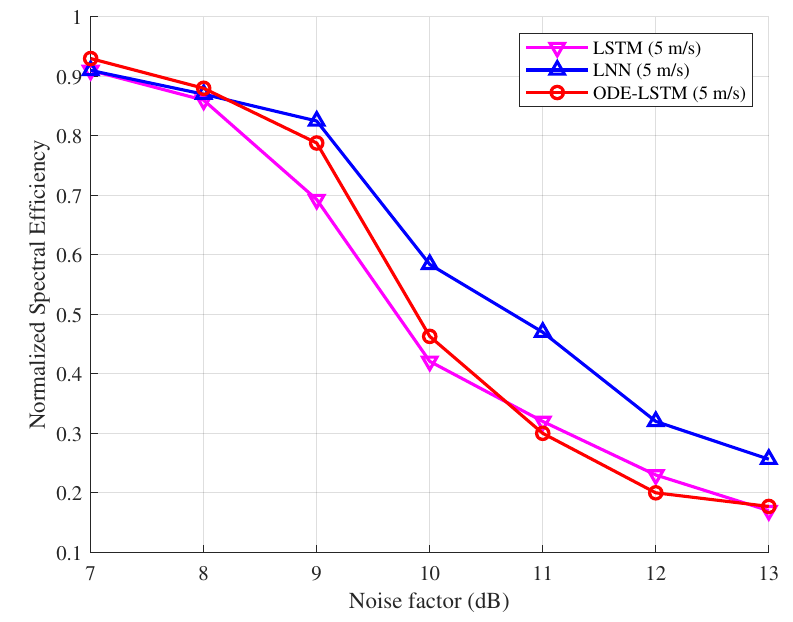}}  
		\vspace{-0mm}
		\captionsetup{name={Fig.}, labelsep=period}
		\caption{Performance evaluation with respect to noise factor.}
		\label{fig:noise_factor} \vspace{-4mm}
	\end{center}
\end{figure}
\vspace{-0mm}
Finally, Fig. \ref{fig:noise_factor} compares the normalized spectral efficiency $\mathrm{SE_N}$ of different schemes in relation to the noise factor $N_\mathrm{F}$, with the UE moving at $v = 5$ m/s. As the noise factor increases, the disturbance in the received signals $\mathbf{y}_n$ intensifies, posing a challenge in learning the data distribution. When $N_\mathrm{F}=7$ dB, ODE-LSTM achieves a $\mathrm{SE_N}$ of 0.93 while LSTM and LNN both achieve a $\mathrm{SE_N}$ of 0.91. However, as $N_\mathrm{F}$ increases, the performance of ODE-LSTM degrades quickly. When $N_\mathrm{F}=11$ dB, ODE-LSTM only attains a $\mathrm{SE_N}$ of 0.30, lower than 0.32 of LSTM and 0.47 of LNN, the proposed scheme achieves 46.9\% higher $\mathrm{SE_N}$ than the LSTM based scheme. In contrast, the performance of the proposed method begins to surpass other schemes when $N_\mathrm{F}$ exceeds 8 dB, demonstrating strong robustness in handling noise interference. The underlying reason can be analyzed as follows: the LSTM employed by ODE-LSTM can handle sequential data but lacks the nonlinear response capability. Conversely, the ODE of ODE-LSTM has the nonlinear response capability, but is limited to creating an interpolation of probability vectors between two time slots. LNN integrates the advantages of LSTM and ODE, offering more nonlinear flexibility in handling sequential data. Consequently, LNN exhibits a more robust response to noise interference.

\section{Conclusion}\label{sec:conclusion}
In order to perform real-time beam tracking in the presence of background noise, a robust continuous-time beam tracking method leveraging liquid neural network was proposed in the paper. This approach allows for dynamic adjustments to the inputs, enhancing adaptability and performance in noisy environments. Simulation results demonstrated that the proposed method can attain up to 46.9\% higher normalized spectral efficiency than the baselines with the UE moving at 5 m/s, which is beneficial to the practical deployment of beam tracking in urban environments with high background noise.  

\section*{Acknowledgement}
The work was supported by the China National Key R\&D Program under Grant 2021YFA1000500 and 2023YFB2904804, National Natural Science Foundation of China under Grant 62331023, 62101492, 62394292 and U20A20158, Zhejiang Provincial Natural Science Foundation of China under Grant LR22F010002, Zhejiang Provincial Science and Technology Plan Project under Grant 2024C01033, and Zhejiang University Global Partnership Fund, MOE Tier 2 (Award number MOE-T2EP50220-0019) and A*STAR (Agency for Science, Technology and Research) Singapore, under Grant No. M22L1b0110.

\bibliographystyle{IEEEtran}
\bibliography{string}
\vspace{12pt}
\end{document}